\documentclass{article}
\title{A HIERARCHY OF DYNAMIC STATES OF RELAXATION}
\author{Ll. Bel \\
Lab. Gravitation et Cosmologie Relativistes. ESA 7065 \\
{\small Tour 22-12, 4 place Jussieu, 75252 Paris}
}
\date{\today}
\begin{document}
\maketitle

\begin{abstract}

We define a hierarchy of dynamic relaxed gas spheres as
solutions of the Poisson equation coupled to a hierarchy of
approximations of the Liouville equation leading, when this equation
is satisfied, to the well-known isothermal gas spheres in the static
case, but also to a new form of dynamic maximum relaxation. The two
previous steps of the hierarchy correspond to an increasing degree of
local relaxation at the center of the configuration. 

\end{abstract}

\section{Introduction}

Embodied into the concept of equilibrium there are two ingredients.
The first and most primitive is the idea of stillness. The second one
is the idea that equilibrium is the latest stage of a macroscopic
process to which a system has been driven by a microscopic mechanism 
at work in every cell of the system. It would be superfluous to
distinguish these two ingredients if we assumed that the process
under consideration necessarily leads to stillness, but we do not
need to do that. Here we distinguish between the concept of equilibrium
proper with
the usual connotation of stillness attached to it, and the concept of
maximum relaxation which is a concept that describes some terminal dynamic
stages of evolution.

We have already considered this distinction
before\,\footnote{Ref. \cite{Bel}} in the framework of
General relativity and more precisely in the description of the
matter contend of some Robertson-Walker models  of universe using
Relativistic kinetic theory. Since the Universe is not stationary,
i.e, there does not exist any global time-like Killing vector-field,
but expanding it is not possible to assume that the
cosmological fluid is in equilibrium in the usual sense. But it is
possible to imagine that this fluid is in an state of maximum
relaxation, which in the quoted reference we called generalized
equilibrium. We believe now that the concept of maximum relaxation
may be useful also in more classical contexts, either to describe
laboratory systems with time-dependent boundary conditions or 
external perturbations, or in astrophysical descriptions of
self-gravitating systems in the strict framework of Newtonian
gravity. This is more specifically the problem that we discuss in this
paper. We hope that this restriction to a more classical problem may
help to uncover the usefulness of the concept of maximum relaxation.

Actually the description in a particular case of an state of
maximum relaxation is not the unique goal that we pursue in this
paper. We are equally interested in describing some intermediate
steps that may settled down in systems which do not reach uniformly maximum
relaxation. It might well be that some of these partial stages of
relaxation become interesting by themselves.

It is possible to distinguish three domains in the physics of the kinetic
theory of
gases: i) the physics of equilibrium, ii) the physics near
equilibrium where perturbation theory can be used, and iii) the
jungle where the most difficult problems arise. This paper is a
modest contribution to a partial clarification of a small territory
of this non perturbative jungle.

\section{Particular field integrals}
 
Let us consider a spherical mass distribution with Newtonian gravitational
potential $V(t,r)$ and a massive particle free-falling  
in the corresponding gravitational field according to the
Newtonian equations of motion:

\begin{equation}
\label {2.1}
\frac{dv^i}{dt}=-\delta^{ij}\partial_j V(t,r), \quad
v^i=\frac{dx^i}{dt}, \quad
r=\sqrt{\epsilon_i(x^i)^2}, \quad \epsilon_i=1 
\end{equation}

We shall say that the following quadratic expression: 

\begin{equation}
\label {2.2}
I=\frac12 A_{ij}v^i v^j+ B_i v^i + C
\end{equation}
is a spherically symmetric {\it Field integral} if:
\begin{itemize}
\item  $A_{ij}$ and $B_i$ have the following tensor structure:

\begin{equation}
\label {2.3a}
A_{ij}=M\delta_{ij}+Nn_in_j, \quad B_i=Pn_i
\end{equation}
with:

\begin{equation}
\label {2.3b}
n_i=\delta_{ij}n^j, \quad n^i=r^{-1}x^i,
\end{equation}
$M$, $N$, $P$ and $C$ being functions of $t$ and $r$. 
\item The total derivative of $I$ with respect to $t$ is independent
of the velocity:

\begin{equation} 
\label {2.3c} 
\frac{dI}{dt}=F(t,r)
\end{equation} 
where $F$ is a function that we shall call the
source of the field integral. \end{itemize} 

Requiring $I$ to satisfy this last equation taking into account
\ref{2.2}, yields, to start with, the following set of conditions:

\begin{equation}
\label {2.3}
\partial_{(k} A_{ij)}=0 
\end{equation}

\begin{equation}
\label {2.4}
\frac12\partial_tA_{ij}+\partial_{(i}B_{j)}=0
\end{equation}

\begin{equation}
\label {2.5}
-\delta^{jk}\partial_k V A_{ij} +\partial_t B_i+\partial_iC=0
\end{equation}

\begin{equation}
\label {2.6}
-\delta^{jk}\partial_j V B_k+\partial_tC=F
\end{equation}
And taking into account \ref{2.3a} we obtain the following system of 
equations: 

\begin{equation}
\label {2.8}
M^\prime+\frac{2N}{r}=0, \quad N^\prime-\frac{2N}{r}=0
\end{equation}

\begin{equation}
\label {2.9}
\dot M+\frac{2P}{r}=0, \quad \dot N+2P^\prime-\frac{2P}{r}=0
\end{equation}

\begin{equation}
\label {2.10}
-V^\prime(M+N)+\dot P+C^\prime=0
\end{equation}

\begin{equation}
\label {2.11}
-V^\prime P+\dot C=F
\end{equation}
where a dot means partial derivative with respect to $t$ and a prime
a partial derivative with respect to $r$.

Integrating the system of Eqs. \ref{2.8}-\ref{2.10} we get:

\begin{equation}
\label {2.12}
N=-\nu r^2, \quad M=\nu r^2+\mu
\end{equation}	

\begin{equation}
\label {2.13}
P=-\frac12 \dot \mu r
\end{equation}
\begin{equation}
\label {2.15}
C=\mu V+\frac14\ddot\mu r^2+\sigma
\end{equation}
where $\nu$ is a constant, and $\mu(t)$ and $\sigma(t)$ are two
functions of $t$.   
Taking into account \ref{2.13} and
\ref{2.15} Eq. \ref{2.11} becomes:

\begin{equation}
\label {2.14}
\dot\mu(V+\frac12 V^\prime r)+\mu\dot V+\frac14{\dot\mu}^{(3)} r^2
+\dot\sigma=F
\end{equation}
where ${\dot\mu}^{(3)}$ means the third derivative of $\mu$ with
respect to $t$.

The concept of a field integral is an obvious generalization of the
concept of constant of motion to which it reduces if $F=0$. This
concept can be useful only if some other considerations restrict its
source $F$. This will be the case in the sequel of this paper.

Collecting the above results, $I$ can be written as:

\begin{equation}
\label{2.16}
I= E+\frac12\nu J^2  
\end{equation}
where:

\begin{equation}
\label{2.17}
E=\mu(\frac12{\vec v}^2+V)-\frac12\dot\mu r\vec v\vec n 
+\frac14\ddot\mu r^2+\sigma, \quad 
J^2=r^2[{\vec v}^2-(\vec v\vec n)^2]. 
\end{equation}
$J$, which is the angular momentum per unit mass with
respect to the center of the mass distribution,
is a constant of the motion, but $E$, which is
proportional to a generalization of the energy per unit mass to which
it reduces if $\mu=1$ and $\dot V=0$ is a field integral:

\begin{equation}
\label{2.17bis}
\frac{dE}{dt}=F, \quad \frac{dJ}{dt}=0 
\end{equation}

\section{Coupling gravitation to a kinetic equation}

We assume in this paper that the density $\rho(t,x^i)$ of the gas
under consideration is given as the mean value	of a kinetic
distribution function $f(t,x^i, v^j)$. More precisely we assume that
the density $\rho$ in the Poisson equation

\begin{equation}
\label {3.1}
\triangle V=\rho
\end{equation}  
where units have been chosen such that:

\begin{equation}
\label {3.1a}
4\pi G=1
\end{equation}
is given by:

\begin{equation}
\label {3.2}
\rho(t,x^i)=\int{f(t,x^i, v^j)\,dv^1dv^2dv^3}
\end{equation}
where $f\ge 0$ is a solution of a kinetic equation:

\begin{equation}
\label {3.3}
\partial_t f+v^i\partial_i f
-\delta^{ij}\partial_i V\frac{\partial f}{\partial v^j}=B
\end{equation}
with $B(t,x^i)$ being a function independent of the velocity that has
to be zero when equilibrium or maximum relaxation is reached.
 
Since we are interested only in spherically symmetric configurations,
$B$, $\rho$ and $V$ will be functions of $t$ and $r$, the equation
\ref{3.1} becoming:

\begin{equation}
\label {3.4}
V^{\prime\prime}+\frac2r V^\prime=\rho, 
\end{equation}
and $f$ will be a function of $t$, $r$, ${\vec v}^2$ and 
$\vec v\vec n$.

Finally, we shall assume that the domain on which \ref{3.2} and \ref{3.3}
apply includes the center of the gas sphere and is bounded 
by a sphere of finite radius $R$. And also that in this domain the density 
and the mass $M$ are finite. 
Consistently with these assumptions we shall assume without any loss
of generality that $V(t,0)=0$, but we have to keep in mind that the
value of this quantity should be chosen appropriately to
match the interior solution to the
solution corresponding to a next layer beyond the surface $r=R$.
Complete models and the matching problems are not considered in this
paper which deals only with the central cores of gas spheres where
presumably some degree of relaxation has already been achieved.

\section{The road to equilibrium or maximum dynamic relaxation}
 
We assume from now on that $f$ is a distribution function that has
the following form:

\begin{equation}
\label {4.4}
f=m\exp(-I)
\end{equation}
where $m$ is the mass of the particles of the gas and 
$I$ is a field integral of the type that we considered in section
2, with $N$, $M$, $P$ and $C$ as in \ref{2.12}, \ref{2.13} and
\ref{2.15}, and with $F$ being given by Eq. \ref{2.14}. 
Therefore the r-h-s of \ref{3.3}
is:

\begin{equation}
\label {4.5ter}
B=-fF
\end{equation} 

Evaluating the integral \ref{3.2} we obtain:

\begin{equation}
\label {4.8}
\rho=(2\pi)^{3/2}m\frac{\exp(-\mu V-\sigma-\frac14\ddot\mu r^2
+\frac18\mu^{-1}{\dot\mu}^2 r^2)}{\mu^{3/2}(1+\mu^{-1}\nu r^2)} 
\end{equation}

We follow below the hierarchy of steps leading to equilibrium or to
maximum relaxation described by an ordered set of restrictive
assumptions on the function $F(t,r)$. We shall consider a power
expansion of the this function with respect to the variable r:

\begin{equation}
\label {4.6}
F(t,r)=F_0(t)+F_1(t)r+\cdots+\frac{1}{n!}F_n(t)r^n+\cdots
\end{equation} 
and we shall examine the models which are obtained assuming that more
and more terms, starting with $F_0$, of this expansion are zero.

Concomitantly with \ref{4.6} we shall consider the power series
expansions of $V(t,r)$ and $\rho(t,r)$ derived from \ref{4.8}:

\begin{equation}
\label {4.9}
V(t,r)=V_0(t)+V_1(t)r+\cdots+\frac{1}{n!}V_n(t)r^n+\cdots
\end{equation}

\begin{equation}
\label {4.10}
\rho(t,r)=\rho_0(t)+\rho_1(t)r+\cdots+\frac{1}{n!}\rho_n(t)r^n+\cdots
\end{equation}

Integrating \ref{3.4} term by term we obtain:

\begin{equation}
\label {4.11}
V_1=0, \quad V_n=\frac{n-1}{n+1}\rho_{n-2}, \quad n>1
\end{equation}
the first of these equalities coming from the assumption that the
density is finite at the center.  

{\it First step}. Assuming that $F_0(t)=0$ in Eq. \ref{2.14} yields:

\begin{equation}
\label {4.12}
\dot\mu V_0+\mu\dot V_0+ \dot\sigma=0, \quad \hbox{or} \quad 
\mu V_0+\sigma=Constant_1,
\end{equation}
and therefore using \ref{4.8} and \ref{4.11} we can write:

\begin{equation}
\label {4.13}
V_2=\frac13\rho_0=a\mu^{-3/2} , \quad \hbox{with} \quad a=Constant_2.
\end{equation} 
From $V_1=0$ it follows that $F_1=0$. Also from \ref{4.8} and $V_1=0$ 
it follows that $\rho_1=0$ and therefore from \ref{4.11} we have also
$V_3=0$. 

{\it Second step}. The next step is then to assume also that $F_2=0$,
i.e.:

\begin{equation}
\label {4.14}
\dot\mu V_2+\frac12\mu\dot V_2+\frac14{\dot\mu}^{(3)}=0
\end{equation}
and using \ref{4.13} we get:

\begin{equation}
\label {4.15}
{\dot\mu}^{(3)}+a\mu^{-3/2}\dot \mu=0
\end{equation}
This is the first relevant result of this paper as it gives the
equation that governs the dynamics of the central core of the spherical
gas configuration. 

This equation has two types of solutions:
\begin{itemize}
\item A. Those with $\dot\mu=0$
\item B. Those with $\dot\mu$ non identically zero	
\end{itemize}
{\it Type A.} In the first case Eq. \ref{2.14} becomes:

\begin{equation}
\label {4.15bis}
\mu\dot V+\dot\sigma=F
\end{equation}
therefore from \ref{4.12} it follows that $F_0=0$. In fact, using
repeatedly Eqs. \ref{4.8}, \ref{4.11} and \ref{4.15bis} it is easy to prove
that $F=0$, meaning that $f$ is a solution of the Liouville equation,
and that $\dot\rho=0$ and $\dot V_n=0$ for $n>0$. The potential
$V(r)$ will be obtained integrating the non-linear differential
equation:

\begin{equation}
\label {4.15a}
V^{\prime\prime}+\frac2r V^\prime= 3a\mu^{-3/2}
\frac{\exp(-\mu V)}{1+\mu^{-1}\nu r^2}
\end{equation} 

If $\nu=0$ then the configurations
thus obtained are the well-known static isothermal gas spheres with
temperature:

\begin{equation}
\label {4.15ter}
T=\frac{m}{k\mu}  
\end{equation}  
where $k$ is Boltzmann's constant. 

{\it Type B1} From now on we
assume that we are dealing with the case B above, i.e. $\dot\mu\not=0$.
In this case the potential $V(t,r)$ is a solution of the following
equation:

\begin{equation}
\label {4.30}
V^{\prime\prime}+\frac2r V^\prime= 3a\mu^{-3/2}
\frac{\exp(-\mu V -(\frac14\ddot\mu-\frac18\mu^{-1}{\dot\mu}^2)r^2)} 
{1+\mu^{-1}\nu r^2}
\end{equation}

{\it Type B2}
Since from \ref{4.11} and $\rho_1=0$ it follows that
$V_3=0$ and therefore also $F_3=0$ the next step consists in
requiring $F_4=0$. This yields:

\begin{equation}
\label {4.16}
3\dot\mu V_4+\mu\dot V_4=0 
\end{equation} 
From \ref{4.11} and \ref{4.8} we have:

\begin{equation}
\label {4.18}
V_4=\frac95a\mu^{-3/2}(-a\mu^{-1/2}-\frac12\ddot\mu+
\frac14\mu^{-1}{\dot\mu}^2 -2\mu^{-1}\nu),
\end{equation}
Deriving this expression and substituting in \ref{4.16} we obtain:

\begin{equation}
\label {4.19}
{\dot\mu}^{(3)}+\frac12\mu^{-1}\dot\mu\ddot\mu+(2a\mu^{-3/2}
-\frac14\mu^{-2}{\dot\mu}^2+2\nu\mu^{-2})\dot\mu=0
\end{equation} 
and using in this equation \ref{4.15} we get at this step the following
equation:

\begin{equation}
\label {4.20}
\ddot\mu=-2a\mu^{-1/2}+\frac12\mu^{-1}{\dot\mu}^2-4\mu^{-1}\nu
\end{equation} 
Going back to \ref{4.18} and \ref{4.11} we see that this
implies that:

\begin{equation}
\label {4.21}
V_4=0, \quad \rho_2=0
\end{equation}
From \ref{4.8} and the preceding results we have $\rho_3=0$ and therefore
also $V_5=0$ and $F_5=0$. 

The equation to integrate to get the potential $V$ is:

\begin{equation}
\label {4.31}
V^{\prime\prime}+\frac2r V^\prime= 3a\mu^{-3/2}
\frac{\exp(-\mu V+(\frac12 a\mu^{-1/2}+\mu^{-1}\nu)r^2)} 
{1+\mu^{-1}\nu)r^2}
\end{equation}  

{\it Type B3}. This will be the last step. From \ref{4.8} and
\ref{4.11} we have:

\begin{equation}
\label {4.23}
V_6=\frac{180}{7}a\nu^2\mu^{-7/2}
\end{equation}
Requiring $F_6=0$ we obtain:

\begin{equation}
\label {4.22}
4\dot\mu V_6+\mu\dot V_6=0 
\end{equation}  
and using \ref{4.23} we get:

\begin{equation}
\label {4.24}
a\mu^{-7/2}\dot\mu\nu=0
\end{equation}
and since we are considering here only dynamic configurations it
follows that:

\begin{equation}
\label {4.24a}
\nu=0
\end{equation}
The evolution equation \ref{4.20} becomes in
this case:

\begin{equation}
\label {4.25}
\ddot\mu+2a\mu^{-1/2}
-\frac12\mu^{-1}{\dot\mu}^2=0
\end{equation}
This is the case of maximum relaxation since, as can easily be
proven,
the results already obtained imply that:

\begin{equation}
\label {4.26}
\rho(t)=\rho_0(t)=3a\mu^{-3/2}, \quad
V(t,r)=V_0(t)+\frac12 a\mu^{-3/2}(t)r^2, \quad F(t,r)=0
\end{equation}   
$V_0(t)$ is at this stage an arbitrary function of $t$ to be fixed by
matching the core solution to its exterior gravitational
field.

\section{Explicit solutions}

We are in this section interested on the explicit solutions of Eqs.
\ref{4.15}, \ref{4.20} and \ref{4.25}, corresponding to the first and
second 
stages of dynamic partial and total relaxation in the domain where
the variable $\mu$ is positive.

Eq. \ref{4.15} has an obvious first integral. Namely:

\begin{equation}
\label {5.1}
b=\ddot\mu -2a\mu^{-1/2}
\end{equation}	
Considering $t$ as a function of $\mu$ an elementary process of
integration leads to:

\begin{equation}
\label {5.6}
\dot\mu=\sqrt{2b\mu+8a\sqrt{\mu}+2c_1}
\end{equation}
from where we obtain
the general solution of \ref{4.15}:

\begin{equation}
\label {5.2}
t-t_0=\int_{\mu(t_0)}^\mu\frac{dy}{\sqrt{2by+8a\sqrt{y}+2c_1}}
\end{equation}
the three constants of integration being $\mu_0$, $b$, and $c_1$
which are
related to $\mu(t_0)$, $\dot\mu(t_0)$ and $\ddot\mu(t_0)$ by the
formulas:

\begin{equation}
\label {5.3}
b=\ddot\mu(t_0) -2a\mu(t_0)^{-1/2}, \quad
c_1=\frac12\dot\mu(t_0)^2-\mu(t_0)\ddot\mu(t_0)-2a\sqrt{\mu(t_0)} 
\end{equation}

The roots of the radical in Eq. \ref{5.6} are:

\begin{equation}
\label {5.4}
{\sqrt{\mu}_\pm}=\frac1b(-2a\pm\sqrt{4a^2-bc_1})
\end{equation} 
and the values of $\ddot\mu$ for these values are:

\begin{equation}
\label {5.5}
\ddot\mu_\pm=\pm\frac{\sqrt{4a^2-bc_1}}{\sqrt{\mu}_\pm}.
\end{equation}
Therefore if ${\sqrt{\mu}_\pm}$ are real and positive then $\mu_+$ 
corresponds to a minimum of the function $\mu$ and
$\mu_-$ corresponds to a maximum.

From the preceding considerations it follows that the solutions of
\ref{4.15} can be classified according to the following types:
\begin{itemize}
\item Type I1: $c_1>0, \ b\geq 0$. The function $\mu$ has neither a maximum

nor a minimum in the domain of interest ($0\leq\mu<\infty$). If the initial
value of $\dot\mu_0$ is positive $\mu$ will increase without limit;
otherwise it will reach the value zero.	
\item Type I2: $c_1>0, \ b<0$. The function $\mu$ has a maximum,
$\mu_-$.
\item Type II1: $c_1=0,\ b\geq 0$. The function $\mu$ has neither a
maximum nor a minimum.
\item Type II2: $c_1=0,\ b<0$. The function $\mu$ has a maximum, $\mu_-$.
\item Type III1: $c_1<0,\ b>0$. The function $\mu$ has a
minimum, $\mu_+$.
\item Type III2: $c_1<0,\ b=0$. The function $\mu$ has a minimum,
$\mu_+$. 
\item Type III3: $c_1<0,\ b<0$. The function $\mu$ 
has a maximum, $\mu_-$, a minimum $\mu_+$, and is periodic; the period
being:

\begin{equation}
\label {5.7}
Period=2\int_{\mu_+}^{\mu_-}\frac{dy}{\sqrt{2by+8a\sqrt{y}+2c_1}}
\end{equation}
\end{itemize}

Let us consider the following function of $t$:

\begin{equation}
\label {5.8}
S(t) \equiv \ddot\mu+2a\mu^{-1/2}-\frac12\mu^{-1}{\dot\mu}^2+4\mu^{-1}\nu
\end{equation}
so that Eq. \ref{4.20} can be written as $S=0$. Evaluating the derivative
of \ref{5.8}
and using Eq. \ref{4.15} we readily get:

\begin{equation}
\label {5.9}
\dot S=-\mu^{-1}\dot\mu S.
\end{equation}
This means that $S=0$ is a conditional first integral of Eq.
\ref{4.15}: if a solution $\mu(t)$ of this equation is such that 
$S(t_0)=0$
for a particular value of $t$ then it is zero for any time.
Equivalently, the solutions of \ref{4.20} are the solutions \ref{4.15}
for which the initial conditions have been constrained to satisfy
\ref{4.20}. 

Using \ref{5.3} and \ref{5.6} in Eq. \ref{4.15} yields:

\begin{equation}
\label {5.10}
c_1=4\nu
\end{equation}
Therefore the general solution of Eq. \ref{4.20} can be described as
the general solution of \ref{4.15} given above requiring $c_1$ to
satisfy \ref{5.10}. In particular from \ref{4.24a} it follows that the
solutions of type II above correspond to configurations of maximum
relaxation. 

\section{Examples}

Let us introduce the temperature function $T(t)$ as the following
obvious generalization of the temperature parameter \ref{4.15ter}:

\begin{equation}
\label {6.1}
T(t)=\frac{m}{k\mu(t)}
\end{equation}
and let us consider a system of units such that besides \ref{3.1a} one has:

\begin{equation}
\label {6.2}
\rho_0(t_0)=3, \quad \mu(t_0)=1
\end{equation}
where $t_0$ is some particular value of $t$. The units of time ($\bar T$), 
length ($\bar L$) and mass ($\bar M$) of this
system measured in the MKS system of units are as follows:

\begin{equation}
\label {6.3}
\bar T=(\frac43\pi G\rho_0(t_0))^{-1/2}\,\hbox{s}, \quad
\bar L=\bar T\mu(t_0)^{-1/2} \,\hbox{m}, \quad
\bar M=(4\pi G)^{-1}{\bar L}^3{\bar T}^{-2} \,\hbox{kg}
\end{equation}

Each set of
values of the parameters $b$, $c_1$ and the total mass $M$ will define a 
three parameter
family of physical examples depending on the assumed physical values
of $\rho_0(t_0)$, $\mu(t_0)$ and $M$ measured in the MKS system of
units. Equivalently $T(t_0)$ can be used if we know $m$. Here we
shall assume that $m$ is the mass of an hydrogen atom.  

We are going to discuss two particular classes of examples belonging
to two different types among those discussed in the preceding section. 

Our first example assumes that:

\begin{equation}
\label {6.4}
b=-2.2,  \quad c_1=-1.8 
\end{equation}
This corresponds to a type III3 and therefore $\mu(t)$ will be a periodic
function. Assuming that $t_0$ is a time corresponding to a maximum value of

$\mu$, then from \ref{4.13} and \ref{5.4} it follows that:

\begin{equation}
\label {6.5}
a=1, \quad \mu_-=1, \quad \mu_+=0.8
\end{equation}
and performing the integral \ref{5.7} we obtain:

\begin{equation}
\label {6.6}
Period=5.4
\end{equation}
From \ref{3.4} we have that the radius $R$ and the mass $M$ of the gas 
configuration at any time are related by the formula:

\begin{equation}
\label {6.7}
M=4\pi R^2(t)V^\prime(t,R(t))
\end{equation} 
the potential function $V(t,r)$ being obtained by integration of Eq.
\ref{4.30}. Assuming, for example, that $M=7.2$ and $\nu=0$ a numerical 
integration yields the following maximum and minimum values of $R$:

\begin{equation}
\label {6.8}
R_{min}=2.5, \quad R_{max}=2.8
\end{equation}	 
The maximum value of the central density corresponding to
the assumptions made above is:

\begin{equation}
\label {6.9}
(\rho_0)_{max}=\rho_0(t_0+Period/2)=5.5
\end{equation}

Remark.- In this example the total mass $M$ compatible with the
values \ref{6.4} has to be less than $\approx 13$.

To see a physical model corresponding to the above parameters 
we have to choose a value of the central density at time
$t_0$ and also a value at the same time for the temperature. Let us
assume as an example that:

\begin{equation}
\label {6.10}
\rho_0(t_0)= 0.5 \, \hbox{kg/m}^3 \quad Temp(t_0)=10^7 \,  \hbox{K}  
\end{equation} 
This corresponds to the following values:

\begin{equation}
\label {6.11}
\bar T=8.5\times 10^4 \, \hbox{s}, \quad
\bar L=4.2\times 10^8 \, \hbox{m}, \quad
\bar M=1.2\times 10^{25} \, \hbox{kg}
\end{equation}  
From \ref{6.6}, \ref{6.8} and \ref{6.9} we have then:

\begin{equation}
\label {6.12}
Period=5.4\times \bar T\, \hbox{s}=4.6\times 10^5 \, \hbox{s} \quad
(\rho_0)_{max}=	 0.9 \, \hbox{kg/m}^3 
\end{equation}
and:

\begin{equation}
\label {6.14}
R_{min}=2.5\times \bar L\ \hbox{m}=1.1\times 10^9 \, \hbox{m}, \quad
R_{max}=2.8\times \bar L\ \hbox{m}=1.2\times 10^9 \, \hbox{m},
\end{equation}

\begin{equation}
\label {6.15}
Mass=7.2\times\bar M=9.0\times 10^{25}\, \hbox{kg}
\end{equation}

Our second example assumes that:

\begin{equation}
\label {6.20}
b=-4, \quad c_1=0
\end{equation}
This corresponds to a case of maximum relaxation of type II2, and
most of its physical behavior is explicit in \ref{4.26} and \ref{5.2}.
In this case also the function $\mu(t)$ has a maximum
(T(t) has minimum). Let us assume that $t_0$ is the value of the time when
this maximum is reached, then  from \ref{5.2} we obtain that the time
interval that it takes for the function $\mu(t)$ to become zero is:

\begin{equation}
\label {6.21}
t_{collapse}=1.1
\end{equation}

If we assume for instance that:

\begin{equation}
\label {6.22}
\rho_0(t_0)= 20\times 10^3 \, \hbox{kg/m}^3
\end{equation}  
we obtain:

\begin{equation}
\label {6.23}
t_{collapse}=465\, \hbox{s}
\end{equation}

\end{document}